\documentclass{article}
\usepackage{spconf,amsmath,graphicx}
\usepackage{tabu}
\usepackage{bm}
\usepackage{amsmath}
\usepackage{amssymb}
\usepackage{cite}
\usepackage{url}
\usepackage{multirow}
\usepackage{graphicx}
\usepackage{tabularx}
\DeclareMathOperator*{\softmax}{softmax}

\ninept
\title{Discovering Sound Concepts and Acoustic Relations In Text}
\name{Anurag Kumar, Bhiksha Raj, Ndapandula Nakashole}
\address{School of Computer Science \\ Carnegie Mellon University\\Pittsburgh, PA, USA - 15213 \\ {\small \tt alnu@andrew.cmu.edu,  bhiksha@cs.cmu.edu,ndapa@cs.cmu.edu} }
\begin{document}
%
\maketitle
\begin{abstract}
In this paper we describe approaches for discovering acoustic concepts and relations in text. The first major goal is to be able to identify text phrases which contain a notion of audibility and can be termed as a sound or an acoustic concept. We also propose a method to define an acoustic scene  through a set of sound concepts. We use pattern matching and parts of speech tags to generate sound concepts from large scale text corpora. We use dependency parsing and LSTM recurrent neural network to predict a set of sound concepts for a given acoustic scene. These methods are not only helpful in creating  an acoustic knowledge base but in the future can also directly help acoustic event and scene detection research.
\end{abstract}
\vspace{-0.1in}
\section{Introduction}
\vspace{-0.1in}
\label{sec:intro}
Analyzing non-speech content has been gaining a lot of attention in the audio community.  
Such non-speech audio content plays an important role in understanding the environment around us. Successful detection of acoustic events and scenes is critical for several applications. One of the most prominent applications is content based retrieval of multimedia recordings \cite{yu2014informedia}\cite{amato2015content}, where the audio component of multimedia carries a significant amount of information. Other well known applications of automated analysis of audio data are: audio based surveillance \cite{atrey2006audio}, human computer interaction \cite{janvier2012sound}, classification of bird species \cite{fagerlund2007bird}, context recognition system \cite{eronen}.

The primary focus in automated machine understanding of non-speech content of audio has been on successful detection and classification of audio events and scenes. Several methods have been proposed for audio event and scene detection \cite{zhuang2010,kumar,foggia2015reliable,kons2013audio,cakir2015polyphonic}. In the most recent DCASE challenge \footnote{\label{dcase} http://www.cs.tut.fi/sgn/arg/dcase2016/}, deep neural network methods dominated performance for audio events whereas factor analysis and Non Negative Matrix Factorization  methods were found to be more promising for acoustic scenes. Moreover, due to the limited availability of labeled data and the time consuming and expensive process of manual annotations, there have been attempts to learn event detectors from weakly labeled data as well \cite{kumar2016audio}\cite{anuragweakly}. These methods rely on weak labels which can be automatically obtained for audio data on web using the associated metadata such as tags, and titles. 

One major limitation in most of the current literature on audio content analysis is the limited vocabulary of audio events. In almost all cases, the analysis is done on a very small set of $5-20$ acoustic events. Clearly, this is very small from for several of the applications we have mentioned. More importantly, mere detection and classification of audio events does not lead to a comprehensive understanding of acoustic concepts. If we look at the analogous problem in the field of computer vision, one can notice that object detection in images has been scaled to thousands of visual object categories \cite{deng2009imagenet}. Moreover, these thousands of categories are  organized into a hierarchical structure which allows higher level semantic analysis. Visual concept ontologies \cite{vco} have been proposed for reducing dependence on text based retrieval of images. The \emph{Never Ending Image Learner} (NEIL) project  \cite{chen_iccv13} not only detects thousands of visual objects and scenes in images but has also learned a variety of commonsense knowledge visual relationships such as \emph{Umbrella looks similar to Ferris wheel} or scene-object relation such as \emph{Monitor is found in Control room}. This allows it to provide visual knowledge to knowledge bases such as  Never Ending Language Learner (NELL)  \cite{carlson2010toward}. Another architecture EventNet \cite{ye2015eventnet} tailored towards multimedia events  organizes $500$ multimedia events using over $4000$ visual concepts. Clearly, these knowledge relations and ontologies are crucial for semantic search of multimedia data on web. 

A similar architecture is desirable for sounds as well which is not only aware of a large number of acoustic concepts but can also draw higher level semantic information and relations about sounds. For example, the system should be aware that \emph{honking}, \emph{beeping} and \emph{engine running} can be related through a common source \emph{car}. Even more important are scene-sounds acoustic relations such as, an acoustic scene \emph{Park} consists of sounds  event \emph{children laughter}. Some works have tried to relate sounds through hand crafted sound taxonomies \cite{salamon2014dataset}\cite{schafer1993soundscape}\cite{brown2011towards}\cite{raimbault2005urban}. Taxonomies for environmental sounds has been of particular interest in these works. Even in the specific context of environmental sounds there is no clear consensus on building such taxonomies \cite{salamon2014dataset}. Different approaches have been applied in different cases and in most cases it is based on subjective opinions. Moreover, in several of these urban taxonomies, a large part of the taxonomy  is made up of broad categories and the number of low-level acoustic concepts is once again very small. This limits their utility for both accumulating sound related knowledge as well as for audio event and scene detection research. In this paper,  we  take a step towards large scale understanding of sound by addressing some of these issues. The motivation is to develop methods which can automatically catalog sounds and generate other commonsense knowledge about sounds. Although not a component of this paper, this can definitely aid in audio event and scene detection tasks.   

First, we try to address the problems of acoustic concept vocabulary by proposing methods for automated discovery of potential sound concepts by applying natural language processing and machine learning techniques on a large corpus of text. For automated discovery of sound concepts we propose  a simple yet effective unsupervised approach based on part of speech (POS) patterns. We follow up on this step by proposing a word embedding based supervised method for classifying a given text phrase into sound phrase (concept) or non sound phrase (non-sound concept). This supervised method allows us to identify the notion of audibility in any given text phrase. Acoustic relations such as acoustic scene- sound concepts, sounds and sources are equally important for understanding sounds. For acoustic relations, we propose a method for automatically describing an acoustic scene or environment through sound concepts. Sound concepts and acoustic scenes are related through dependency paths and then an LSTM neural network \cite{hochreiter1997long} is used to predict whether a sound concept is found in an acoustic scene or not. Although, in this work we looked into the specific case of \emph{scene-concept} relations, our method can be extended to other forms of relations as well, for example \emph{concept-source} relations. Scene-concept relations can be extremely helpful in creating sound ontologies. To the best of our knowledge, this is the first work on text based understanding of sounds, leading to large scale acoustic knowledge. The rest of the paper is as follows.

In Section 2 we describe our methods for discovering sound concepts, in Section 3 we describe the dependency path and LSTM based approach for \emph{scene-concept} relations. We describe our subjective and objective evaluation for the proposed methods in Section 4 and finally we conclude in Section 5. 
\vspace{-0.15in}
\section{Sound Concepts}
\vspace{-0.1in}
\label{sec:sc}
Automated discovery of text phrases which can be designated as sound concepts is a very difficult problem. First,  whether a text phrase has a notion of sound or audibility in it can be very subjective and dependent on the context in which it is used. There  are unigram and bigram phrases such as \emph{music, laughter, glass breaking, woman screaming } etc. which on its own gives an impression of sound or audibility in it. However, in several cases a direct belief of sound might not be apparent from the text phrase on its own but it is either a source of or is directly related to salient acoustic phenomena which is well understood in commonsense human knowledge. Examples include phrases such as \emph{helicopter, birds, dog, car engine} . These acoustic concepts appear in several audio event databases and it is expected to have capabilities to detect these events. Hence, any large list of sound concepts should include such phrases. Automated discovery of sound concepts (phrases) becomes difficult due to this subjective and contextual way of expressing sound concepts. We propose a simple yet very effective method of obtaining ``audible" text phrases or sound concepts from large text corpora. We also propose a supervised method of classifying a given text phrase as sound concept or non-sound concept by incorporating syntactic and semantic content of phrase through word embeddings. 
  
\vspace{-0.15in}
\subsection{Unsupervised Sound Concept Discovery}
\vspace{-0.1in}
We introduce an unsupervised method for discovering sound concepts in text. Our method is based on the idea that there are patterns in specific forms that are primarily used to express sound concepts in language. These patterns can help in identifying sound concepts. 

We begin with a single pattern: ``sound(s) of \textless Y\textgreater"  where  \textit{Y} is any phrase, we allow $Y$ to be up to  4 words long. We  then  look for occurrences of the pattern in a large corpus. In our experiments, we used the English part of  ClueWeb09\footnote{http://lemurproject.org/clueweb09/}
From this we obtain  a large collection of  occurrences such as: ``\text{sound} \text{ of\textit{ honking cars}}", ``\text{sound} \text{ of  \textit{gunshots}}".

However, this step  produces a significant amount of noise.  We therefore treat its output as  \textit{candidate sound concepts} and  introduce a minimally-supervised method for  pruning noise from this collection.  First, we generalize candidate concepts by replacing  mentions with their part of speech tags, as follows:
\begin{align*}
		\text{ sound} \text{ of honking cars} =  \text{ sound} \text{ of VBG  NN}  \\
		\text{sound } \text{ of  gunshots} = \text{ sound} \text{ of NNS}  
\end{align*}
where the part of speech (POS) tag   $VBG$  denotes verbs in the gerund form, $NN$, and $NNS$ denote singular and plural nouns, respectively \footnote{POS Abbreviations: https://www.ling.upenn.edu/ courses/Fall\_2003/ling001/penn\_treebank\_pos.html}. The  POS generalized concepts reduce the data size to about  only $20$ unique patterns. Since the POS patterns are so few, we can  use them to filter out noisy concepts with little effort. The key to filtering is that not all POS patterns express valid concepts.  We can eliminate all but $6$   of the POS patterns. For example, the pattern ``sound of JJ (adjective)" does not  express  sound concepts. All candidate concepts that match the $6$ valid POS patterns are retained  and the rest are discarded. The full list of valid POS patterns  with examples is shown in Table \ref{tab:patterns}. The patterns in Table \ref{tab:patterns} produced a total of $116,729$ unique sound mentions from the corpus. 
\begin {table}[t]
	\centering
	\begin{tabu}{|ll|l|}
		\hline
		\multicolumn{1}{|c}{}& \textbf{Pattern} & \textbf{ Example Concept}\\
		\hline
		\multicolumn{1}{|c|}{P1} & \textless \textit{X}\textgreater\hspace{0.1cm}of\hspace{0.1cm} (DT) VBG NN(S)  & honking cars \\
		\multicolumn{1}{|c|}{P2}  & \textless\textit{X}\textgreater \hspace{0.1cm}of\hspace{0.1cm} VBG	&  yelling  \\
		\multicolumn{1}{|c|}{P3}  & \textless \textit{X}\textgreater \hspace{0.1cm}of\hspace{0.1cm} (DT) NN(S) VBG & dogs barking  \\
		\multicolumn{1}{|c|}{P4} & \textless \textit{X}\textgreater \hspace{0.1cm}of\hspace{0.1cm} (DT) NN(S) &  gunshots  \\
		\multicolumn{1}{|c|}{P5}  &  \textless \textit{X}\textgreater \hspace{0.1cm}of\hspace{0.1cm} (DT) NN NN(S) &   string quartet \\
		\multicolumn{1}{|c|}{P6} & \textless \textit{X}\textgreater \hspace{0.1cm}of\hspace{0.1cm} (DT) JJ NN(S) &  classical music   \\
		\hline
	\end{tabu}
	\caption{Patterns for  discovering sound concepts in text.   $VBG$ is the part of speech tag  for verbs in the gerund form,  $NN$  for  singular nouns (S means plular),  $DT$ for  determiners, and $JJ$ for adjectives.}
	\label{tab:patterns}
	\vspace{-0.25in}
\end {table}
\vspace{-0.15in}
\subsection{Supervised Classification}
\label{sec:supclas}
\vspace{-0.1in}
The unsupervised discovery of sound phrases (concepts) in the previous section can still contain non-sound phrases. A few examples of such phrases which are clearly not sound concepts but do not get filtered out by the two step process in the  previous section are \emph{someone being (NN VBG), price dropping (NN VBG), gaining experience (VBG NN), happy hunters (JJ NNS)}. Hence, to improve upon the unsupervised discovery of potential sound concepts, we propose a supervised method for classifying a text phrase as sound phrase (sound concept) or non sound phrase (non sound concept). 

Since bigram phrases are the most dominant and expressive set of sound concepts discovered by the unsupervised method, we focus specifically on bigram phrases. A set of labeled data is required for supervised training of classifiers for text phrase classification. To obtain a completely reliable set of labeled data, we manually inspect a small subset of the sound concepts obtained in the previous section and mark if it is actually a sound concept or not. Note that, in the unsupervised case only $6$ POS patterns express valid sound concepts. We use  the rest of the POS patterns to create a list of  negative examples. We manually inspect and label a small subset of this list as well. Finally, we end up with a total of $\sim 6000$, sound concept and non-sound concept phrases.  

The text phrases need to be appropriately represented by features on which classifiers can be trained. \emph{Word Embeddings} have been found to be very effective in capturing syntactic and semantic similarity between words \cite{mikolov2013distributed,pennington2014glove,hashimoto2015task} and have shown remarkable success in a variety of semantic tasks \cite{baroni2014don}. Word embeddings map words into a fixed dimensional vector representation. In this work we use \emph{word2vec} \cite{mikolov2013distributed} to obtain vector representation for words in text phrases. We use Google News pre-trained embeddings \footnote{\label{word2vec} https://code.google.com/archive/p/word2vec/} to represent each word by $300$ dimensional vectors. We then use two methods for representing each bigram phrase. In the first case, we take the average of the word2vec representation for each word to represent the whole phrase. This representation is referred to as \emph{AWV}. In the second case, we concatenate the vectors (\emph{CWV}) for each word to obtain a $600$ dimensional vector for each phrase. These vectors can then be used for training any classifier such as Support Vector Machine (SVM) to build a sound vs non-sound phrase classifier.   
\vspace{-0.15in}
\section{Acoustic Relations}
\vspace{-0.1in}
\label{sec:sr}
In this section we describe an approach to learn relationships in the domain of acoustic world or \emph{acoustic relations}. Acoustic relations can be of different forms such as \emph{acoustic scene - concepts } relations: sound concepts found in an acoustic scene, \emph{source-sound} relations: source of the corresponding sound, co-occurrence relations:  sounds which often occur together. In this paper we focus specifically on \emph{scene-concept} relations where our goal is to describe an acoustic scene or environment by a set of sounds which occur in that scene or environment. Information in the  form of what types of sounds make up a scene can be extremely helpful in audio scene recognition tasks$^1$. Moreover, these relations also provide co-occurrence information about sound concepts. For example, \emph{laughing} and \emph{cheering} often occur together in several acoustic environment. These additional information about sound concepts can be exploited in a variety of applications. From the perspective of semantic analysis in text, we cast this task as a relation classification problem.
\begin{table}[t]
\centering
\caption{36 Acoustic environments used in experiments }
\label{tab:envs}
\resizebox*{1.0\columnwidth}{!}{
\begin{tabular}{|c|c|c|c|c|}
\hline  
& \multicolumn{4}{c|}{Acoustic Environments} \\ 
\hline
1 & Office & Farm & House & Bus \\
2 & Parties & Funeral & Library & Park \\
3 & Street & Parking Lot & Church & Train \\
4 & Airplane & Wedding & Cafe & Cities \\
5 & Campus & Ballgame & Bathroom & Classroom \\
6 & Train Station & School & Parks & Bar \\
7 & Grocery Store & Trucks & Forest & Restaurant \\
8 & Subway & Airport & Arena & Construction \\
9 & Beach & Garden & Stadium  & Ranch \\
\hline
 \end{tabular}
}
\vspace{-0.15in}
\end{table}

First we find all sentences in the ClueWeb corpus that mention at least one of the $116,729$ sound concepts discovered in Section \ref{sec:sc}, and at least one acoustic environment such as  ``beach", ``park", etc.  In our experiments, we worked with  a total of $36$ acoustic environments which we define in Table \ref{tab:envs}, but our method is generic and can work with any number of environments.  Most of acoustic scenes from DCASE $^1$ scene classification challenge are part of our setup as well. We then apply a dependency parser\footnote{https://pypi.python.org/pypi/practnlptools/1.0}  to  any sentence that mentions a sound concept and an acoustic environment. This step produces   dependencies that form a directed graph, with words being nodes and dependencies being edges. For example, the sentence: \textit{``The park was filled with the sound of children playing"} , yields the following dependencies:
\begin{center}
	\begin{minipage}[c]{0.8\linewidth}
		\textit{det(park-2, The-1)}\\
		\textit{nsubjpass(filled-4, park-2)}\\
		\textit{auxpass(filled-4, was-3)}\\
		\textit{root(ROOT-0, filled-4)}\\
		\textit{det(sound-7, the-6)} \\
		\textit{nsubj(playing-10, sound-7)}\\
		\textit{prep\_of(sound-7, children-9)}\\
		\textit{prepc\_with(filled-4, playing-10)'}
	\end{minipage}
\end{center}
	
The details of the dependency relations can be found in \cite{de2008stanford}. Next, we traverse the dependency graph in order to obtain the path between the mention of a sound concept, in this case ``children playing", and the mention of the acoustic environment ``park". Shortest paths between entities have been found to be a good indicator of relationships between entities \cite{XuMLCPJ15,NakasholeWS13}. We therefore extract the shortest path. In our example, the shortest path  labeled with edge and node names  is as follows: ``nsubjpass() filled prepc\_with() sound prep\_of()".
\vspace{-0.1in}
\subsection{Training Data} 
\vspace{-0.1in}
Given the paths, we would like to classify scene-sound pairs into those that express the relationship of interest  (SoundFoundInEnvironment) and those that do not. Classifier training would require labeled training data.  

To obtain training data, we proceed as follows:  We sort the paths by frequency, that is, how often we have seen the path occur with different scene-sound   pairs. Among the most frequent paths, we label the paths yes or no, depending on whether they express the  relationship of interest. This gives us  a way to generate positive and negative examples using the labeled paths.
Examples of paths that generate positive training data are shown in Table \ref{tab:pospts}.  Examples of   paths that generate  negative  training data are shown in Table \ref{tab:negpts}. 

\begin{table}[t]
\centering
\caption{Example Paths for Positive Training Data}
\label{tab:pospts}
\resizebox*{1.0\columnwidth}{!}{
\begin{tabular}{|c|c|}
\hline  
\textit{prep\_along()} & \textit{ prep\_of() sound nsubjpass() heard prep\_in()} \\
\textit{prep\_of()} & \textit{nsubjpass() filled prep\_with() sound prep\_of()} \\
\textit{prep\_of() sound prep\_on()} & \textit{conj\_and() sounds prep\_of()} \\
\textit{prep\_with() sounds prep\_of()} & \textit{  prep\_of() sound prep\_to()}\\
\textit{nsubj() alive prep\_with() sound prep\_of()} & \textit{ prep\_upon()} \\
\textit{ prep\_of() sounds prep\_from()} & \textit{ prep\_of() sounds prep\_on()} \\
\textit{  prep\_of() sound nsubj() came prep\_from()} & \textit{prep\_of() sounds prep\_at()} \\
\hline
 \end{tabular}
}
\vspace{-0.20in}
\end{table}

\begin{table}[t]
\centering
\caption{Example Paths for Negative Training Data}
\label{tab:negpts}
\begin{tabular}{|c|c|}
\hline  
\textit{conj\_and()} & \textit{ amod()} \\
\textit{poss()} & \textit{nn()} \\
\textit{nn() sound prep\_of()} & \textit{ prep\_through()} \\
\textit{prep\_of()} & \textit{appos()}\\
\textit{det()} & \textit{conj\_and() sound prep\_of()}\\
 \textit{prep\_to()} & \textit{prep\_of() sound nsubj() filled dobj() } \\
\hline
 \end{tabular}
\vspace{-0.20in}
\end{table}
\vspace{-0.1in}
\subsection{Classification} 
\vspace{-0.1in}
We use an LSTM  recurrent  neural network to learn the  scene-sound relationship. Each word $w$ is  mapped to a  $d$-dimensional  vector $\bm{v_w} \in \mathbb{R}^d$ through an embedding matrix $\bm{E} \in \mathbb{R}^{ |V| \times d} $, where $|V|$ is the vocabulary size, and each row corresponds to a vector of a  word.  We initialize the word embeddings with the 300-dimensional Google News pre-trained embeddings$^4$. For the dependency relation names ``amod()", since they do not have entries in the pre-trained embedding matrix, we randomly initialize their vector embeddings, and learn them during training.
	
	\noindent \textbf{Path Encoding.}   To encode the  shortest path between a sound concept and an acoustic scene, we use  an LSTM  recurrent  neural networks (RNN) which is capable of  learning long range dependencies. While regular  RNNs can also  learn long dependencies, they tend be biased towards recent inputs in the sequence. LSTMs  tackle  this limitation with   a memory cell  and an adaptive  gating mechanism that controls how much  of the input to give to the memory cell, and the how much of   the previous state to forget \cite{hochreiter1997long}.
	
	 We have a  path: $\bm{p} =\bm{p}_1, ..., \bm{p}_p \in  \mathbb{R}^d$  and an associated path matrix $\bm{P} \in \mathbb{R}^{p \times d}$,  where each row corresponds to the embedding vector of the word in that position.
	
	\noindent The LSTM mention encoder  generates  the path encoding, $\bm{v_p}$,  as follows:
	\begin{eqnarray}\label{lstmencoder}
	\bm{h}_i &=& LSTM(\bm{v}_{p_i}, \bm{h}_{i-1}, \bm{c}_{i-1}),  i = 1, \ldots, p \\
	\bm{v}_p &=& \bm{h}_i : {i =p} \nonumber
	\end{eqnarray}
	The LSTM  encodes the word  at timestep $i=t$ in the path using the word embedding vector $\bm{v}_{p_t}$,  the previous output $\bm{h}_{t-1}$, and the previous state of the LSTM cell $\bm{c_}{t-1}$. The output $\bm{h}_t$   is computed using the four main elements in the LSTM cell: an input gate $\bm{i}_t$, a forget gate $\bm{f}_t$, an output gate $\bm{o}_t$, a
	memory cell $\bm{c}_t$ with a self-recurrent connection. The cell takes as input a $d$-dimensional input vector for  word $\bm{x}_t = \bm{p}_i$, the previous hidden state $\bm{h}_{t-1}$, and the memory cell $\bm{c}_{t-1}$. It  calculates the new vectors using the following equations:
	\vspace{-0.15cm}\begin{eqnarray}
	\bm{i}_t&=&\sigma\left(\bm{W}_{xi}  \bm{x}_t+\bm{U}_{hi} h_{t-1} + \bm{b}_i\right),\\
	\bm{f}_t&=&\sigma\left(\bm{W}_{xf} \bm{x}_t+\bm{U}_{hf} \bm{h}_{t-1} + \bm{b}_f\right),\nonumber\\
	\bm{o}_t&=&\sigma\left(\bm{W}_{xo} \bm{x}_t+\bm{U}_{ho}  \bm{h}_{t-1} + \bm{b}_o\right),\nonumber\\
	\bm{u}_t&=&\tanh\left(\bm{W}_{xu} \bm{x}_t+\bm{U}_{hu} \bm{h}_{t-1} + \bm{b}_u\right),\nonumber\\
	\bm{c}_t&=&\bm{i}_t {\odot} \bm{u}_t + \bm{f}_t {\odot} \bm{c}_{t-1},\nonumber\\
	\bm{h}_t&=&\bm{o}_t {\odot} \tanh(\bm{c}_t),\nonumber
	\end{eqnarray}
	where $\sigma$ is the sigmoid function, $\odot$ is element-wise multiplication, the  $W$ and $U$ parameters  are weight matrices, and the $b$ parameters are bias vectors.
	
	\noindent{\textbf{Prediction}.} From path encoding  $\bm{v}_p$,  we  compute the output of the neural network, a distribution over the positive and negative labels.  The output for each path is decoded by a linear layer and a  \textit{softmax} layer  into probabilities over the  two labels.   Therefore, the prediction  $\bm{d}_r$
\begin{equation}
\label{mention_model}
\bm{d}_r = \softmax(\bm{W}_r \cdot \bm{v}_p )
\end{equation}		
where 	$\softmax(z_i) = e^{z_i}/\sum_{j}e^{z_j}$.
\vspace{-0.15in}
\section{Analysis and Evaluation}
\vspace{-0.1in}
\begin {table}[t]
\centering
\caption{Analysis of Unsupervised Sound Concept Discovery}
\label{tab:unsupres}
\resizebox*{1.0\columnwidth}{!}{
\begin{tabular}{|c|c|c|c|}
\hline
& Pattern & $\#$ Concept & $+$ in Top 100 Freq. \\
\hline
P1 & \textless \textit{X}\textgreater\hspace{0.1cm}of\hspace{0.1cm} (DT) VBG NN(S)  & 9335 & 98\\
P2  & \textless \textit{X}\textgreater \hspace{0.1cm}of\hspace{0.1cm} VBG	&  1395 & 71 \\
P3  & \textless \textit{X}\textgreater \hspace{0.1cm}of\hspace{0.1cm} (DT) NN(S) VBG & 19194 & 91 \\
P4 & \textless \textit{X}\textgreater \hspace{0.1cm}of\hspace{0.1cm} (DT) NN(S) & 20064 & 59\\
P5  &  \textless \textit{X}\textgreater \hspace{0.1cm}of\hspace{0.1cm} (DT) NN NN(S) & 26473 & 93 \\
P6 & \textless \textit{X}\textgreater \hspace{0.1cm}of\hspace{0.1cm} (DT) JJ NN(S) &  40268 &  49\\
\hline
\end{tabular}
}
\vspace{-0.2in}
\end {table}
In this Section we analyze and evaluate our proposed methods. The complete list of sound concepts discovered by our method is available on this \cite{soexpt} webpage. We had six unique POS patterns for filtering candidate sound concepts. The total number of sound concepts corresponding to each POS pattern is shown in Table \ref{tab:unsupres}. Since, the total number of sound concepts discovered is fairly large, manually inspecting it to identify actual sound concepts among the discovered ones is very difficult. For each concept discovered by our method we maintain a count of total number of times that sound concept occurred in the text corpus. We select the \emph{Top 100} for manual inspection and identify concepts which can actually be labeled as a \emph{sound concept}. We do it for each POS pattern. The number of positive ($+$) hits in this Top 100 most frequent concepts is shown in Table \ref{tab:unsupres}. We note that $3$ POS pattern have more than $90$ positives. The lowest is for \textless JJ NN(S) \textgreater. However, note that the frequency of occurrence of a discovered concept has nothing to do with it being \emph{truly} a sound concept. Discovered concepts such as \emph{sobbing voices, siren breaking, cheering crewmen}  contain an impression of sound and are positive examples of sound concepts but occur very few times in the text corpus. Hence, the purpose of column $2$ in Table \ref{tab:unsupres} is to show how well our method did on phrases which occurred frequently in our process.   

As described in Section \ref{sec:supclas}, we created a list of sound concepts (positive) and non-sound concepts (negative) bigram phrases. The total number of positive examples is $3189$ and total number of negative examples is $2758$. We randomly divide this data into $4$ folds. $3$ folds are used for training and then the trained model is tested on left out fold. The experiment is done all 4 ways. Linear SVMs are trained on both \emph{AWV} features and \emph{CWV} features. The accuracies for both feature representation are shown in Table \ref{tab:supres}. Concatenated word2vec features gives slightly better performance compared to averaged word2vec features. An average accuracy of more than $90\%$ is achieved which shows that our supervised classifier is highly reliable in classifying a text phrase as sound or non-sound phrase.    
\begin {table}[t]
\centering
\caption{Accuracy of Supervised Classification}
\label{tab:supres}
\begin{tabular}{|c|c|c|c|c|c|}
\hline
& Fold 1 & Fold 2 & Fold 3 & Fold 4 & Avg \\
\hline
AWV & 87.03 & 89.05 & 87.84 & 89.77 & 88.42 \\
\hline
CWV & 90.00 & 89.32 & 91.87 & 90.30 & 90.37 \\
\hline
\end{tabular}
\vspace{-0.1in}
\end {table}
\begin {table}[t]
\centering
\caption{Examples of Environment (Scene)-sounds relations discovered by our method}
\label{tab:relations}
\resizebox*{1.0\columnwidth}{!}{
\begin{tabu}{|l|l|}
	\hline
	\textbf{Environment} & \textbf{Sounds}	\\
	\hline
	Forest & Birds Singing, Breaking Twigs, Cooing, Falling Water	\\
	\hline
	Restaurant & Jazz, Laughter, People Talking, Music Drifting \\
	\hline	
	Airport & Planes Flying, Plane Engines, Aircraft, Intercoms \\
	\hline
	Park & Laughing, Police Siren, Birds Chirping, Footsteps \\
	\hline	
	Ranch & Horses, Gunfire, Tapping Water, Bulldozers \\
	\hline	
	Church & Children Laughing, Church Bells, Singing, Applause\\
	\hline	
	Beach & Waves Crashing, Waves Lapping, Surf Hitting\\
	\hline
	Construction & Hammering, Jackhammers, Engines, Blasting\\
	\hline
	Street & Sirens, Men Shouting, Honking Cars, Cheering \\
	\hline
	Bar & Piano Playing, Laughter, Clinking Glasses, Cheering \\
	\hline
\end{tabu}
}
\vspace{-0.2in}
\end {table}

Table \ref{tab:relations} shows a few examples of \emph{scene-concept} relations found by the system. Some unusual findings are \emph{Rifle Shots} in \emph{Library}, \emph{Chirping Birds} in \emph{Library}. The full list of sound concepts discovered for each acoustic scene or environment is available on this \cite{soexpt} webpage. A subjective analysis of all discovered relations shows that for most of the relations discovered are meaningful in the sense that the sound concept is actually found in that acoustic environment.  

\vspace{-0.1in}
\section{Conclusions}
\vspace{-0.1in}
In this paper we presented methods for text based understanding of sounds and acoustic relations. We proposed a method for automated discovery of sound concepts using a large text corpus. It discovered over $100,000$ sound concepts and to the best of our knowledge no such other exhaustive list of sound concepts exists in current literature. We found among the discovered concepts, those corresponding to POS patterns in form of \textless VBG NN(S) \textgreater and  \textless NN(S) VBG \textgreater are in general very reliable. This is clearly expected as a large number of sounds are related to its sources through some action. We also proposed a simple word embedding based method for learning supervised classification of text phrases in sound or non-sound phrases (concepts). This supervised method achieved an accuracy of over $90\%$. Although, the total number of examples considered in supervised classification experiments is not very large ($\sim 6000$), it does validate our proposed word embedding based approach. An important aspect of any knowledge base about sounds would be to relate different sounds. In this work we took on the specific case of scene-concept relations where we try to find out the sound concepts which may occur in an acoustic scene or environment. This is helpful in defining an acoustic scene by the sounds which occur in that scene and hence can be exploited in acoustic scene recognition tasks. Other meta level inferences can also be drawn from such acoustic relations. We continue to investigate into methods for discovering other sound related knowledge using text. 

\vfill\pagebreak

\bibliographystyle{IEEEbib}
\bibliography{refs}

\end{document}